\documentclass[prl,reprint,superscriptaddress,showpacs,showkeys,longbibliography]{revtex4-1}
\usepackage{amsmath,amsfonts,amssymb,bm,graphicx,color}

\usepackage[colorlinks=true,linkcolor=blue,urlcolor=blue,citecolor=blue]{hyperref}

\makeatletter
\newcommand*{\balancecolsandclearpage}{%
  \close@column@grid
  \cleardoublepage
  \twocolumngrid
}
\makeatother

\begin{document}

\title{Signatures of a Quantum Griffiths Phase close to \\ an Electronic Nematic Quantum Phase Transition}

\author{Pascal Reiss}
\email[]{pascal.reiss@physics.ox.ac.uk}
\affiliation{Clarendon Laboratory, Department of Physics, University of Oxford, Oxford, UK}

\author{David Graf}
\affiliation{National High Magnetic Field Laboratory, Florida State University, Tallahassee, FL, USA}

\author{Amir A. Haghighirad}
\affiliation{Clarendon Laboratory, Department of Physics, University of Oxford, Oxford, UK}
\affiliation{Institute for Quantum Materials and Technologies, Karlsruhe Institute of Technology, Karlsruhe, Germany}

\author{Thomas Vojta}
\affiliation{Department of Physics, Missouri University of Science and Technology, Rolla, MO, USA}

\author{Amalia I. Coldea}
\email[]{amalia.coldea@physics.ox.ac.uk}
\affiliation{Clarendon Laboratory, Department of Physics, University of Oxford, Oxford, UK}

\date{\today}

\begin{abstract}
 In the vicinity of a quantum critical point, quenched disorder can lead to a quantum Griffiths phase, 
accompanied by an exotic power-law scaling with a continuously varying dynamical exponent that diverges in the zero-temperature limit.
Here, we investigate a nematic quantum critical point in the iron-based superconductor
FeSe$_{0.89}$S$_{0.11}$ using applied hydrostatic pressure. We report an
unusual crossing of the magnetoresistivity isotherms in the non-superconducting
normal state which features a continuously varying
dynamical exponent over a large temperature range. We interpret our
results in terms of a quantum Griffiths phase caused by nematic islands that
result from the local distribution of Se and S atoms. At low temperatures,
the Griffiths phase is masked by the emergence of a Fermi
liquid phase due to a strong nematoelastic coupling and a Lifshitz
transition that changes the topology of the Fermi surface.
\end{abstract}

\pacs{}


\maketitle

\footnotetext[1]{See Supplemental Material at [URL will be inserted by publisher] for further experimental data and analyses}

\paragraph{Introduction}
A central characteristic of finite- and zero-temperature phase transitions is how the spatial and temporal correlation lengths evolve as the transition is approached. For clean and continuous phase transitions, scaling theory predicts power-law divergences of both correlation lengths as a function of control parameter, with the critical exponents reflecting the universality class. Moreover, the spatial and temporal correlation lengths are closely related by the dynamics of the system \cite{Hertz1975,Hertz1976,Millis1993,Moriya1985}. In the presence of quenched disorder, this relation may be lost. Quenched disorder is perfectly correlated in time, but can harbor a spatially varying order parameter. In this situation, a smeared phase transition can occur, where ordered islands form within a disordered bulk \cite{Kuo2016,Cui2018}. Moreover, when order parameter fluctuations within the islands are non-negligible, a Griffiths phase can emerge which leads to continuously varying critical exponents as a function of temperature and control parameter, fundamentally different to clean systems \cite{Griffiths1969,Fisher1992,Fisher1995,DelMaestro2008,Vojta2003,Vojta2006,Vojta2010}.

\begin{figure}[h!]
	\includegraphics[trim={0cm 0cm 0cm 0cm}, width=1\linewidth,clip=true]{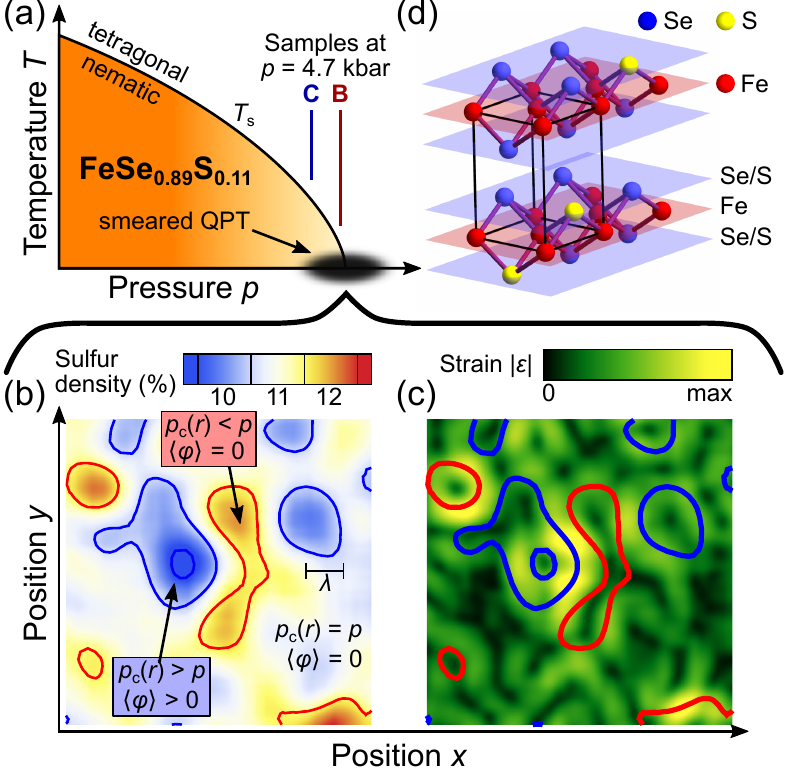}
	\caption{(a) Pressure-temperature phase diagram and (d) crystal structure of FeSe$_{0.89}$S$_{0.11}$. The relative position of samples B and C under a pressure of $p = 4.7$\,kbar are indicated by vertical lines. (b,c) The spatial distribution of small S atoms induces locally varying critical pressures $p_c(r)$ and random local strains. Close to the nematic quantum phase transition (QPT), this leads to the formation of nematic islands. The scale shows the experimental mean free path length $\lambda$ \cite{Reiss2019}.}
	\label{fig:summary}
\end{figure}

Experimentally, quantum Griffiths phases have been identified in ferromagnetic nickel vanadium alloys \cite{Ubaid2010,Wang2017}
as well as a range of superconducting thin films
\cite{Seidler1992,Schneider2012,Shi2014,Xing2015,Saito2018,Lewellyn2019,Liu2019a,Lin2015}.
In the latter systems, sharp crossings of the magnetoresistivity isotherms emerged as a distinctive experimental signature. A scaling analysis revealed a tem\-per\-a\-ture-de\-pen\-dent critical exponent $z\nu$ that diverges in the low-temperature limit. This is the hallmark of a quantum Griffiths phase ($\nu$ is the critical correlation length exponent, and $z$ the dynamical exponent) \cite{Fisher1992,Fisher1995,DelMaestro2008,Vojta2003,Vojta2006,Vojta2010}.

In this Letter, we report the magnetoresistivity of the quasi-2D bulk superconductor FeSe$_{0.89}$S$_{0.11}$ when tuned to the vicinity of its nematic quantum critical point (QCP) using a hydrostatic pressure of $4.7$\,kbar (Fig.~\ref{fig:summary}(a)) \cite{Reiss2019}.
Here, the magnetoresistivity isotherms show a remarkably sharp crossing at about $30$\,T over
nearly two decades in temperature up to $30$\,K.
Scaling of the magnetoresistivity yields a critical exponent $z\nu$
that diverges at low temperatures, in agreement with the quantum Griffiths scenario.
We argue that the Griffiths phase is induced 
by the local distribution of isoelectronic Se and S atoms 
that promote the formation of nematic islands in the vicinity of the nematic QCP, as shown in Fig.~\ref{fig:summary}(b) and (c).
Below a crossover temperature $T \approx 10$\,K, the quantum Griffiths phase and the associated QCP appear to be masked by an emergent non-zero energy scale which coincides with the re-entrance of Fermi liquid behavior attributed to a strong nematoelastic coupling, as well as a topological Lifshitz transition of the Fermi surface.

\paragraph{Methods}
Single crystal of FeSe$_{1-x}$S$_x$ with $x = 0.11$ sulfur substitution were grown using the KCl/AlCl$_3$ chemical vapour transport method as described elsewhere \cite{Bohmer2016}. High-pressure, high-field measurements for samples B and C were carried out in the 45\,T hybrid DC facility in Tallahassee. We used Daphne Oil 7575 as pressure medium which ensures hydrostatic conditions for much higher pressures than reported here, and we used the Ruby fluorescence shifts below 4\,K to determine the pressure. Low-field measurements up to 13.5\,T were carried out on sample A in a QuantumDesign PPMS in Oxford. Here, Daphne Oil 7373 was used, and the pressure was determined by the superconducting transition of tin. Samples were aligned with the magnetic field parallel to the crystallographic $c$ axis to avoid breaking an in-plane symmetry. Transport measurements were performed using a standard 4 or 5 contact setup, using the AC LockIn technique with a low frequency $f \approx 20$\,Hz, and a low excitation current $I_p = 1$\,mA within the $(ab)$ plane.

\paragraph{Results}

\begin{figure}[h!]
\includegraphics[trim={0cm 0cm 0cm 0cm}, width=1\linewidth,clip=true]{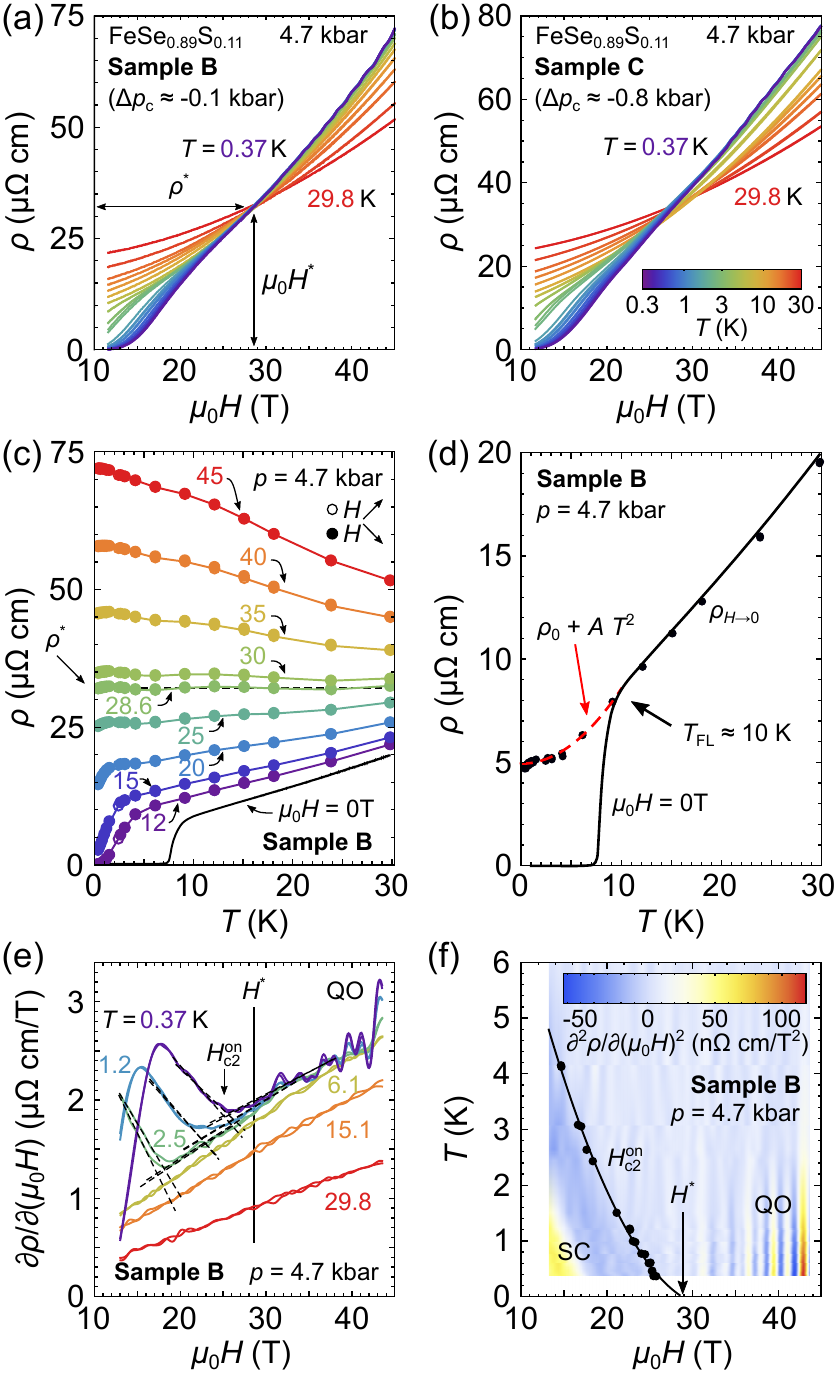}
		\caption{
		(a,b) The isothermal magnetoresistivity of samples B and C at a pressure of $p = 4.7$\,kbar cross at $\mu_0 H^\star \approx 28$\,T, $\rho^\star \approx 33\,\mu\Omega$\,cm. Up and down sweeps show no hysteresis. The pressure difference in brackets represents the distance to the critical pressure (see Fig.~\ref{fig:summary}(a) and in the SM \cite{Note1}).
		(c) The same data as in panel (a), but as a function of temperature in fixed field.
		(d) The actual and extrapolated zero-field resistivities, $\rho(T)$ (solid line) and $\rho_{H \to 0}$ (points, see the SM \cite{Note1}). Error bars are smaller than the symbol size. The red dashed line is a fit to Fermi liquid behavior ($\rho_0 \approx 4.9\,\mu\Omega$\,cm, $A \approx 0.036\,\mu\Omega\,\text{cm}/\text{K}^2$).
		(e) The first derivative reveals the onset of superconductivity, indicated by arrows. Large quantum oscillations (QO) can be seen for $H > H^\star$.
		(f) The extrapolation of the superconducting (SC) onset coincides with $H^\star$ only at $T = 0$.
		All reported data are measured at $p = 4.7$\,kbar.}
	\label{fig:data}
\end{figure}

Figure~\ref{fig:data}(a) and (b) show the temperature dependence of the magnetoresistivity of two different single crystals B and C of FeSe$_{0.89}$S$_{0.11}$  under a hydrostatic pressure of $p = 4.7$\,kbar, which is in the immediate vicinity of their nematic QCPs
($p_c = 4.8$\,kbar for sample B and $5.5$\,kbar for sample C, respectively 
as shown in Fig.~\ref{fig:summary}(a) and in the Supplemental Material (SM) ~\cite{Note1,Reiss2019,Xiang2017}).
All magnetoresistivity isotherms cross around a similar magnetic field,
 $\mu_0 H^\star \approx 28.6$\,T  for sample B and $28.0$\,T  for sample C,
with similar resistivities $\rho^\star \approx 32\,\mu\Omega$\,cm and $34\,\mu\Omega$\,cm, respectively.
This crossing occurs over nearly two decades in temperature $0.3\,\text{K} \lesssim T \lesssim 30\,\text{K}$
and its significance becomes evident in the resistivity plots as a function of temperature in constant field, shown in Fig.~\ref{fig:data}(c).
For $H < H^\star$, the resistivity follows a metallic-like behavior with $\partial \rho / \partial T > 0$ before the sample
becomes superconducting below $T_c^\text{on} \approx 10$\,K (Fig.~\ref{fig:data}(d)).
Equivalently, the onset magnetic field, $H_{c2}^\text{on}$, between the superconducting and normal phases can be identified in magnetic fields
smaller than $H^\star$ in Fig.~\ref{fig:data}(e) whose zero-temperature extrapolation coincides with $H^\star$ (Fig.~\ref{fig:data}(f)).
Thus, the magnetoresistivity crossing occurs strictly within the non-superconducting normal phase for all finite temperatures
 ($H^\star > H_{c2}^\text{on}(T)$), implying that this behavior
 describes the normal phase in the vicinity of the nematic QCP.

In high magnetic fields above $H^\star$,
the resistivity shows insulating-like behavior ($\partial \rho / \partial T < 0$),
 before it saturates below $T \approx 2$\,K (Fig.~\ref{fig:data}(c)), similar to previous reports \cite{Bristow2019}.
Despite this insulating-like behavior, the large magnetoresistivity
is a feature of the metallic, compensated multi-band system FeSe$_{1-x}$S$_x$ \cite{Watson2015b,Coldea2019,Bristow2019}.
Quantum oscillations are visible for temperatures below $\approx 5$\,K (Fig.~\ref{fig:data}(e)),
demonstrating the existence of a Fermi surface and highlighting the high quality of the samples \cite{Reiss2019,Shoenberg1984}.
A two-band analysis of the magnetoresistivity allows us to extrapolate
the zero-field resistivity from high magnetic fields \cite{Note1},
which indicate Fermi liquid behavior below a crossover temperature $T_\text{FL} \approx 10$\,K,
shown in Fig.~\ref{fig:data}(d) \cite{Reiss2019,Bristow2019}.
The orbitally averaged effective masses from quantum oscillations
show non-divergent electronic correlations in the vicinity of the nematic QCP, as discussed in detail in Ref.~\citenum{Reiss2019},
likely due to a coupling between the nematic order parameter and the lattice  \cite{Paul2017,Wang2019,DeCarvalho2019,Vieira2019}.

Next, we use a prototypical power-law scaling ansatz to describe the magnetoresistivity of FeSe$_{0.89}$S$_{0.11}$, previously applied in thin-film materials, including dirty films of FeSe
 \cite{Seidler1992,Schneider2012,Shi2014,Xing2015,Saito2018,Lewellyn2019,Liu2019a,Lin2015}.
In $d$ dimensions, the scaling is given by
\begin{align}
\label{eq:scaling}
\rho(H,T) / \rho^\star = T^{(2-d)/z} f\left(\mu_0|H-H^\star| / T^{1/z\nu}\right)
\end{align}
with $f(0) = 1$ and the critical exponent $z\nu$ \cite{Fisher1990}.
Clearly, a crossing of the magnetoresistivity isotherms at a finite $\rho^\star$
is only possible for a two-dimensional system.
Indeed, FeSe$_{1-x}$S$_x$ have strongly two-dimensional electronic and superconducting properties \cite{Terashima2014,Reiss2019,Coldea2019,Bristow2020Hc2,Farrar2020,Coldea2020review}.

\begin{figure}[htbp]
\includegraphics[trim={0cm 0cm 0cm 0cm}, width=1\linewidth,clip=true]{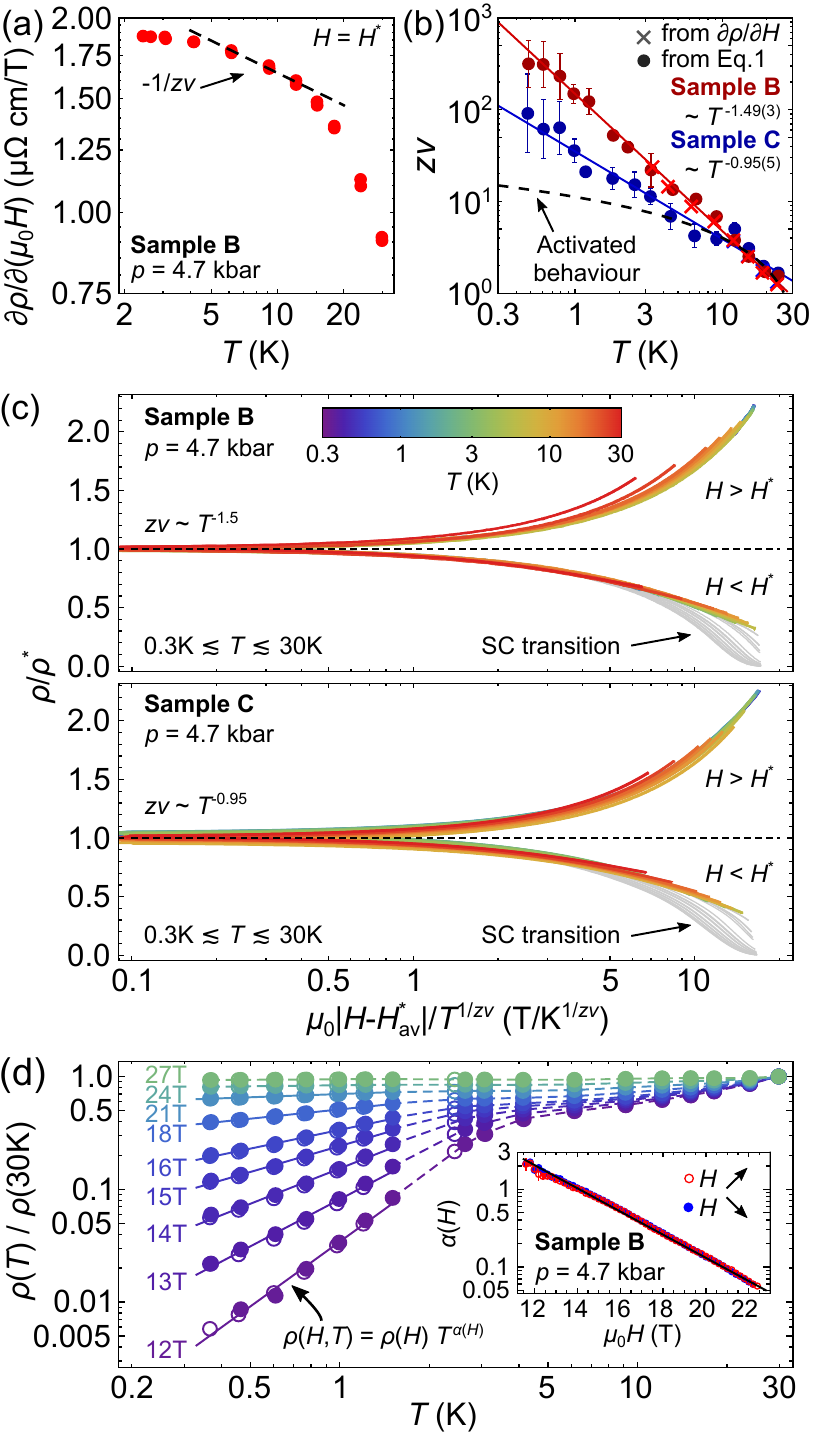}
	\caption{(a) Log-log plot of the first derivative at the crossing field
$H^\star$ (refer to Fig.~\ref{fig:data}(e)). Error bars are smaller than the symbol size. The slope of the dashed line corresponds to $1/z\nu(T)$.
             (b) Temperature dependence of $z\nu$ extracted from panel A (crosses), and from the piece-wise extraction shown in the SM (dots) \cite{Note1}. Error bars indicate a 1$\sigma$ confidence interval.
             (c) Scaled magnetotransport data using $z\nu \sim T^{-1.5}$ (sample B) and $z\nu \sim T^{-0.95}$ (sample C). The superconducting transitions (SC) deviate from this scaling form.
             (d) The low-field, low-temperature resistivity in the mixed state follows a power-law form.
             All reported data are measured at $p = 4.7$\,kbar.}
	\label{fig:scaling}
\end{figure}

For a clean QCP, $z\nu$ is a constant given by the
appropriate universality class, and as described in the SM~\cite{Note1}, this should lead to a straight line in Fig.~\ref{fig:scaling}(a).
This is evidently not the case here where $z\nu$ varies as a function of temperature, shown in Fig.~\ref{fig:scaling}(b).
 Using this extracted $z\nu(T)$,
all magnetoresistivity data collapse onto a single curve for both samples,
reflecting the form of the scaling function $f$, shown in Fig.~\ref{fig:scaling}(c).
Deviations for this scaling only occur for
the superconducting transition at lowest fields and temperatures,
and at the highest temperatures and fields. 
These deviations indicate the limits of the scaling relation, as shown in the SM \cite{Note1}.

\begin{figure*}[htbp]
\includegraphics[trim={0cm 0cm 0cm 0cm}, width=1\linewidth,clip=true]{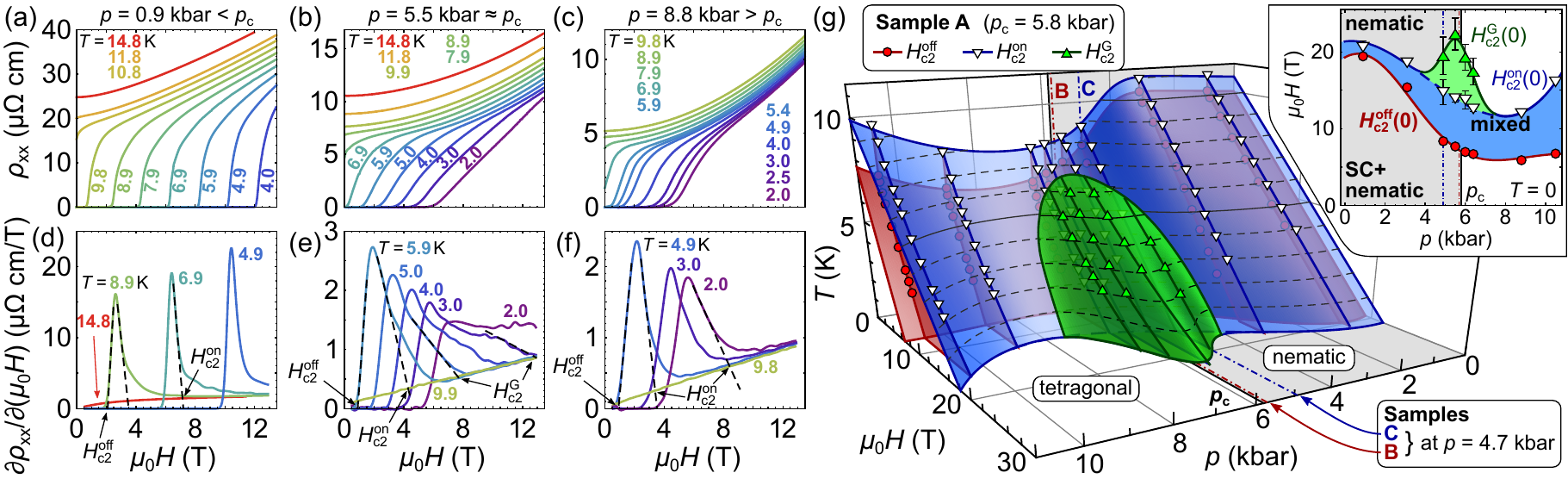}
	\caption{Evolution of superconductivity for sample A.
	(a)-(c) Magnetoresistivity and (d)-(f) the corresponding first derivatives showing the development of the superconducting transition in magnetic fields for pressures across the nematic quantum phase transition.
	(g) Three-dimensional $H$-$p$-$T$ superconducting phase diagram. The inset shows the extrapolated critical fields, $H_{c2}^\text{off}$, $H_{c2}^\text{on}$ and $H_{c2}^\text{G}$, defined in panels (d)-(f), in the zero-temperature limit. 
The relative positions of samples B and C under a pressure of 4.7 kbar in the phase diagram is indicated by dashed and the dotted lines, respectively 
(see also Figs.~\ref{fig:data} and \ref{fig:scaling}).
 Error bars indicate a 1$\sigma$ confidence interval.}
	\label{fig:superconductivity}
\end{figure*}

This scaling analysis reveals an interesting and unexpected feature. While the zero-temperature divergence of the effective critical exponent $z \nu(T)$ is a
key signature of quantum Griffiths phases, the observed power-law form
$z \nu(T) \sim T^\alpha$ (with non-universal exponents
$\alpha \approx -1.5$ for sample B and $\approx -1.0$ for sample C) is much stronger than the
logarithmic (`activated') dependence 
predicted within the infinite-randomness criticality scenario \cite{Vojta2010,Lewellyn2019,Note1}, as shown in Fig.~\ref{fig:scaling}(b).
In fact, a power-law divergence of $z\nu(T)$ is incompatible
with the notion of a normal QCP because the temperature
term $T^{1/z\nu(T)}$ 
in Eq.~\ref{eq:scaling}
remains finite for $T \to 0$, which implies the persistence of a non-zero energy scale at
lowest temperatures.
Interestingly, we find that $z\nu$ deviates from the activated behavior dependence below $T \approx 5$--$10$\,K
which coincides with a re-entrance of Fermi liquid behavior, Fig.~\ref{fig:data}(b). This suggests a suppression of Griffiths fluctuations and/or dimensional changes, e.g. topological changes of the Fermi surface \cite{Reiss2019,Bristow2020,Coldea2019,Coldea2020review}.

We now focus on the nature of the underlying phases separated by $H^\star$.
Figure~\ref{fig:scaling}(d) shows that for fields smaller than $H^\star$,
the resistivity within the superconducting mixed state follows a power-law form $\rho \propto T^{\alpha(H)}$ over almost one decade in temperature. We attribute this power-law form to a disordered vortex-liquid phase that freezes into
a vortex-glass in the zero-temperature limit, as found in underdoped cuprates \cite{Shi2014}.
Crossing over into the high-field regime above $H^\star$ where quantum oscillations are present,
the resistivity reflects the behavior of a metallic phase with a partial charge-carrier localization,
as discussed in the SM \cite{Note1}.

To elucidate the origin and extend of the low-field disordered vortex phase,
we investigate the pressure dependence of the superconducting to normal transitions in magnetic fields on sample A ($p_c \approx 5.8$\,kbar \cite{Reiss2019}).
Figure~\ref{fig:superconductivity}(a)-(f) shows the magnetoresistivity and its derivative up to $13.5$\,T inside the nematic phase ($0.9$\,kbar),
close to the nematic quantum phase transition ($5.5$\,kbar) and within the tetragonal phase ($8.8$\,kbar).
In the nematic and tetragonal phases, the normal-to-superconducting transition widths are nearly temperature and field independent.
In contrast, a visible broadening of the transition is found close to $p_c$,
but only for high fields/low temperatures, thus coinciding with the vortex-liquid phase in sample B.
To quantify this additional broadening,
we extract the superconducting offset and onset critical fields, $H_{c2}^\text{off}$ and $H_{c2}^\text{on}$, as shown in Fig.~\ref{fig:superconductivity}(d)-(f).
Furthermore, we define a critical magnetic field $H_{c2}^\text{G}$,
where the magnetoresistivity derivative has an additional shoulder before it returns to its high-temperature normal state background,
which is observable only in the vicinity of $p_c$
(Fig.~\ref{fig:superconductivity}(e)).
Figure~\ref{fig:superconductivity}(g) summarizes all extracted critical fields and their zero-temperature extrapolations, see also the SM \cite{Note1}.
Interestingly, the zero-tem\-perature superconducting transition width peaks
at the nematic quantum phase transition, doubling the extent of the superconducting mixed state.
The width of the $H_{c2}^\text{G}$(0) peak in pressure is estimated to be $\sigma_p \approx 0.7(2)$\,kbar,
which agrees well with an estimate for the pressure range
of the quantum Griffiths phase, as discussed below.
Figure~\ref{fig:superconductivity}(g) also shows that the zero-field superconducting
transition does not display any similar broadening. This demonstrates that the peak in $H_{c2}^\text{G}$
is a low-temperature and high-field effect, ruling out effects of possible pressure inhomogeneities, as discussed in the SM \cite{Note1}.

\paragraph{Discussion}

Quantum Griffiths phases were previously detected in inhomogeneous superconductor-to-in\-su\-la\-tor transitions
in thin films, including FeSe \cite{Seidler1992,Schneider2012,Shi2014,Xing2015,Saito2018,Lewellyn2019,Liu2019a,Lin2015}.
Here, in bulk FeSe$_{0.89}$S$_{0.11}$, the situation is very different.
The scaling relation only describes the normal state resistivity and holds for magnetic fields up to $45$\,T and
 temperatures up to $30$\,K, vastly exceeding the bulk superconducting phase.
We therefore propose that the quantum Griffiths phase in FeSe$_{0.89}$S$_{0.11}$
emerges from the suppression of the nematic phase with pressure \cite{Reiss2019,Xiang2017,Matsuura2017}
and the formation
of rare nematic islands in a tetragonal matrix
due to the random distribution of sulfur atoms (Fig.~\ref{fig:summary}), as suggested before (see the SM to Ref.~\citenum{Kuo2016}).
To demonstrate how this can lead to a quantum Griffiths phase, we sample a random distribution of $11$\,\% S atoms over a square lattice,
and average the effective sulfur density $x(r)$ over the experimental quasiparticle mean-free path length $\lambda \approx 350$\,\AA\ \cite{Reiss2019},
as shown in Fig.~\ref{fig:summary}(b) and (c).
The intrinsic local variation $\Delta x(r) \approx 0.4\%$ (std.~dev.) gives rise to regions with higher (lower) S content
which have a locally lower (higher) critical pressure $p_c(r)$.
This is the prototypical case of random-mass disorder that smears the quantum phase transition over a region $\Delta p_c(r)$.
By comparing the reported nematic transition temperatures from pressure and isoelectronic
substitution studies, we estimate $\Delta x(r) \propto \Delta p_c(r) \approx 0.4$\,kbar \cite{Coldea2019,Bristow2019,Reiss2017,Xiang2017}.
This estimate is similar to the observed pressure range of a broadened superconducting transition, $\sigma_p = 0.7(2)$\,kbar. This suggests
that the peak in $H_{c2}^\text{G}(0)$ occurs either due to enhanced superconducting
fluctuations within the nematic islands, and/or superconducting nematic islands below the percolation threshold, which get suppressed at $H^\star$.
These effects
could also provide a favorable environment for the observed inhomogeneous superconducting vortex phase in the vicinity of the nematic QCP.
Finally, we note that the spatial arrangement of the S atoms locally breaks the $C_4$ symmetry
of the lattice and thus introduces random-field effects. In the two-dimensional
regime, they may limit the size of the nematic domains, but for weak disorder,
the corresponding breakup length is exponentially large \cite{Seppala2001}.

It is rather surprising that in such a clean system signatures
of a quantum Griffiths phase are detected
in the vicinity of the electronic nematic quantum phase transition.
It is clear that there are a further ingredients at play here
as the nematoelastic coupling sets in and quenches the two-dimensional quantum critical nematic fluctuations,
below a cross-over temperature $T_\text{FL} \approx 10$\,K.
As a result, Fermi liquid behavior with finite electronic correlations is restored \cite{Reiss2019,Bristow2020,Coldea2019,Coldea2020review},
and the quantum Griffiths phase is cut off, leading to the overly strong divergence of $z\nu$.
Additionally, a band with likely 3D character is formed due to a Lifshitz transition of the Fermi surface in the proximity of the nematic QCP \cite{Coldea2019,Reiss2019} which may change the effective dimensionality of the system at low temperatures.

The observation of a quantum Griffiths phase in an iron-based superconductor
has a number of important implications and provides new insights into
the nature of nematic quantum phase transitions.
Our study provides evidence that the quantum Griffiths phase significantly affects the mixed state of the superconducting phase.
Moreover, it may be present in other families of iron-based superconductors in which nematic and tetragonal
phases may form over limited compositional ranges around QCPs \cite{Bohmer2015,Hosoi2016,Bristow2020,Coldea2019}.
Finally, we uncover a number of new experimental signatures in electronic transport measurements,
most notably the reported power-law behavior of $z\nu(T)$,
that could provide new insights into the dynamics of (quenched) nematic order parameter fluctuations.
We hope that our results will guide further theoretical and experimental research in understanding nematic quantum Griffiths phases.

\paragraph{Acknowledgements}
We thank R.~Fernandes for insightful discussions.
This work was mainly supported by the EPSRC (EP/I004475/1, EP/I017836/1).
P.R.~and A.A.H. acknowledge financial support of the Oxford Quantum Materials Platform Grant (EP/M020517/1).
A portion of this work was performed at the National High Magnetic Field Laboratory, supported by NSF Cooperative Agreement DMR-1157490 and the State of Florida.
Work in  Missouri was supported by the NSF (DMR-1828489).
We acknowledge financial support of Oxford University John Fell Fund.
A.I.C.~acknowledges an EPSRC Career Acceleration Fellowship (EP/I004475/1) and
Oxford Centre for Applied Superconductivity for financial support.

\end{document}